\documentclass[11pt]{article}
\renewcommand{\theequation}{\thesection.\arabic{equation}}

\usepackage{amsfonts}
\usepackage{amsbsy}

\newcommand{\BE}{\begin{equation}}
\newcommand{\EE}{\end{equation}}
\newcommand{\BA}{\begin{eqnarray}}
\newcommand{\EA}{\end{eqnarray}}

\newcommand{\at}{\tilde{a}}
\newcommand{\OO}{\langle {\cal O} \rangle}
\newcommand{\gammaO}{ \gamma_{{\scriptscriptstyle\cal O}}}

\begin{document}

\begin{titlepage}

\vspace*{17mm}
\begin{center}
               {\Large\bf Optimization for factorized quantities} 
\\
\vspace*{2mm}
               {\Large\bf in perturbative QCD}
\vspace{22mm}\\
{\Large P. M. Stevenson}
\vspace{7mm}\\
{\large\it
T.W. Bonner Laboratory, Department of Physics and Astronomy,\\
Rice University, Houston, TX 77251, USA}
\vspace{30mm}\\
{\bf Abstract:}

\end{center}

    Perturbative calculations of factorized physical quantities, such as moments 
of structure functions, suffer from renormalization- and factorization-scheme dependence.  
The application of the principle of minimal sensitivity to ``optimize'' the scheme choices is 
reconsidered, correcting deficiencies in the earlier literature.   The proper 
scheme variables, RG equations, and invariants are identified.  Earlier results of Nakkagawa and 
Ni\'egawa are recovered, even though their starting point is, at best, unnecessarily 
complicated.  In particular, the optimized coefficients of the coefficient function $C$ are shown 
to vanish, so that $C^{\rm opt}=1$.  The resulting simplifications mean that the optimization 
procedure is as simple as that for purely-perturbative physical quantities.

\vspace*{1mm}

\end{titlepage}

\setcounter{equation}{0}

\section{Introduction}

The application of the principle of minimal sensitivity \cite{OPT} to the problem of 
factorization-scheme dependence has had a rather unfortunate history.  
The present author shares some of the blame, and this paper aims to make amends.  
The pioneering work by Politzer \cite{Politzer}, which showed the way, was marred by a 
trivial algebraic error, seemingly showing that the optimization equations had no solution.  
The error was belatedly corrected in Ref.~\cite{StePol}.  However, Ref.~\cite{StePol} 
is, in retrospect, insufficiently general beyond second order.  The formulation of 
Nakkagawa and Ni\'egawa (NN) in a series of papers \cite{NN1}-\cite{NN3} is, at best, unnecessarily 
complicated and creates spurious difficulties.  However, NN's optimization equations are 
actually equivalent to those we derive below.   We discuss their work in Appendix A.    
Note that in Refs.~\cite{Politzer}-\cite{NN3} ``$b$'' has the opposite sign to ours.\footnote{
Our notation follows Ref.~\cite{OPTnew}, except that we now omit tilde's on $\Lambda$ 
and $\rho_j$, which had merely emphasized a difference in definition from previous conventions.  
Tildes will be needed here for another purpose.}

   The prototypical factorization problem is in deep-inelastic leptoproduction, 
where a high-energy lepton collides with a proton, or other hadron, 
exchanging a virtual photon of large virtuality $Q^2$. 
Neglecting power-suppressed terms, the $n$th moment, 
$\int_0^1 \! \frac{dx}{x} \, x^n F(x,Q)$,    
of the non-singlet 
proton structure function can be factorized into the form
\BE
\label{Fndef}
F_n(Q) = \langle {\cal O}_n(M) \rangle \,  C_n(Q, M),
\EE
where $ \langle {\cal O}_n(M) \rangle$ is an operator matrix element, $C_n$ 
is a coefficient function, and $M$ is some arbitrary ``factorization scale.'' 
(From now on the moment index $n$ will be suppressed.)

     The operator matrix element $\langle {\cal O}(M) \rangle$ has an 
$M$ dependence given by its anomalous dimension
\BE
\label{gamOdef}
\frac{M}{\OO} \frac{d \OO}{d M}
\equiv \gammaO .
\EE
While $ \langle {\cal O}(M) \rangle$ itself cannot be calculated perturbatively, its 
anomalous dimension, $\gammaO$, has a calculable perturbation series of the form
\BE
\label{gamOser}
\gammaO(a) =- b g a(1 + g_1 a + g_2 a^2 + \ldots).  
\EE
The leading-order coefficient is written as $-b g$ for later convenience.  While $g$ is 
invariant the other coefficients, $g_1, g_2, \ldots$ are scheme-dependent.  
The expansion parameter, $a=a(M)$, is the couplant in some arbitrary 
renormalization scheme (RS) with renormalization scale $M$.  Its $M$ 
dependence is given by the $\beta$ function:  
\BE
\label{betaMeq}
M \frac{\partial a}{\partial M} = \beta(a) = - b a^2 (1+ c a + c_2 a^2 + \ldots).
\EE 
The scheme-dependent coefficients $c_2, \ldots$ can be regarded as RS 
labels \cite{OPT,OPTnew}.

     The coefficient function $C$ can be calculated as a perturbation series:
\BE
\label{Coeffser}
C(Q,M)= 
1 + r_1 \at+ r_2 \at^2 + \ldots,
\EE
where $\at$ is the couplant of some other arbitrary RS -- which can be different from 
the RS used to define $a$.   It can have a different renormalization scale $\tilde{M}$, 
and different RS labels $\tilde{c}_2,\ldots$.  (In the latter respect we differ from 
Ref.~\cite{StePol}.)
Perhaps the easiest way to understand that the RS's for $a$ and $\at$ can be distinct, 
without inconsistency, is to imagine that first both $\OO$ and $C$ are calculated in the 
same RS and then a substitution $\at = a(1+ v_1 a+ v_2 a^2 + \ldots)$, with arbitrary
$v_1, v_2,\ldots$, is made in the result for $C$.  In terms of renormalization constants, 
the $Z_{{\scriptscriptstyle\cal O}}$ constant needed for the renormalization of the operator 
${\cal O}$ (which is genuinely an infinite change of normalization) must be consistent between 
the calculations of $C$ and $\gamma_{{\scriptscriptstyle\cal O}}$, but the 
reparametrization step -- the substitution of $ a= Z_a a_{\rm bare}$ and 
$ \at= \tilde{Z_a} a_{\rm bare}$ in the bare forms of $\gammaO$ and 
$C$, respectively -- can involve distinct 
$Z_a$ and $\tilde{Z_a}$ renormalization constants.

    Thus, what we shall call ``RS/FS dependence'' involves a choice of factorization 
scheme (FS), parametrized by $g_1, g_2, \ldots$, and two, independent, choices of 
RS for $a$ and $\at$ that are labelled, respectively, by 
$\tau$, $c_2, c_3, \ldots$ and by 
$\tilde{\tau}$, $\tilde{c}_2, \tilde{c}_3, \ldots$, 
where
\BE
\tau \equiv b \ln(M/\Lambda), \quad\quad \tilde{\tau} \equiv b \ln(\tilde{M}/\Lambda).
\EE
(See Appendix~B for the definition of $\Lambda$.  Without loss of generality, we may 
assume that the two renormalization prescriptions 
for $a$ and $\at$ are defined so that their $\Lambda$ parameters are the same.)
%
%

         Integrating Eq.~(\ref{gamOdef}), utilizing the $\beta$-function equation, 
gives 
\BE
\OO = ({\mbox{\rm const.}}) 
\exp \left( \int^{a} dx \frac{\gammaO (x)}{\beta(x)} \right).
\EE
Note that the $M$ dependence of $\OO$ comes solely from $a$ (whereas the $M$ 
dependence of $C$ comes solely from the $r_i$ coefficients).  
The constant of integration may be written as a constant $A$ defined by
\BE
\label{Adef}
\OO = A  \exp \left( \int_0^{a} dx 
\frac{\gammaO(x)}{\beta(x)} - 
\int_0^\infty dx \frac{g x}{x^2(1+c x)} \right),
\EE
where, as with the definition of $\Lambda$, the lower limit of $x \to 0$ 
in each integral produces a divergence that cancels between the two integrals.   
The normalization constant $A$ is not calculable from perturbation theory, but is 
RS/FS invariant, as shown in Ref~\cite{StePol}.

\section{Second-order approximation}
\setcounter{equation}{0}

   We first discuss second order, where all authors are in agreement.  
   A second-order approximation corresponds to truncating the series  
for $\gammaO$, $C$, and $\beta$ after two terms.
The integrals in Eq.~(\ref{Adef}) become
\BA
& &  \int_0^{a} dx 
\frac{-b g x(1+g_1 x)}{-b x^2(1+c x)} - 
\int_0^\infty dx \frac{g x}{x^2(1+c x)}
\nonumber \\
& = &
g g_1 \int_0^{a} dx \frac{1}{1+c x} - g \int_{a}^{\infty} dx 
\left( \frac{1}{x} - \frac{c}{1+c x} \right) 
\nonumber
\\
& = & g \left( \frac{g_1}{c} \ln(1+ c a) + \ln(c a) - \ln(1+ c a) \right),
\EA
which exponentiates to
\BE
\label{OO1}
(c a)^g (1+c a)^{-g(1-g_1/c)}.
\EE
Substituting in Eq.~(\ref{Fndef}), one obtains the second-order 
approximation to $F$ as 
\BE
\label{F2ndord}
F^{(2)} = A ( c a)^g (1+c a)^{-g(1-g_1/c)} (1+ r_1 \at).
\EE
This approximant depends on RS/FS choices through three variables, 
$\tau$, $\tilde{\tau}$, and $g_1$.  
Partial differentiations of Eq.~(\ref{F2ndord}) yield
\BA
\label{F2taut}
\frac{1}{F^{(2)}} \frac{\partial F^{(2)}}{\partial \tilde{\tau}} & = & 
\frac{1}{(1+r_1 \at)} \left(  - \at^2 (1+ c \at)  r_1 + 
\at \, \frac{\partial r_1}{\partial \tilde{\tau}} \right),
\\
\label{F2tau}
\frac{1}{F^{(2)}} \frac{\partial F^{(2)}}{\partial \tau} & = & 
- g a (1+g_1 a) + \frac{\at}{(1+r_1 \at)} \,  \frac{\partial r_1}{\partial \tau},
\\
\label{g1feq}
\frac{1}{F^{(2)}} \frac{\partial F^{(2)}}{\partial g_1} & = & 
\frac{g}{c} \ln(1+ c a)+ \frac{\at}{(1+r_1 \at)} \,  \frac{\partial r_1}{\partial g_1}. 
\EA
Self-consistency of perturbation theory requires these variations to be of order $a^2$.  
Noting that $\at=a(1+O(a))$, we see that
\BE
\label{SP15}
\frac{\partial r_1}{\partial \tilde{\tau}} =0, \quad
\frac{\partial r_1}{\partial \tau} = g, \quad 
\frac{\partial r_1}{\partial g_1} =-g, 
\EE
so that $r_1$ has the form
\BE 
\label{SP16}
r_1 = g \left( \tau - g_1 - \boldsymbol{\sigma}_1(Q) \right),
\EE 
where $\boldsymbol{\sigma}_1(Q)$ is an invariant.\footnote{
The earlier literature is a bit sloppy at this point, as we discuss 
in section 4.
} 

      Substituting Eq.~(\ref{SP15}) back into Eqs.~(\ref{F2taut}--\ref{g1feq}) 
and equating to zero produces the optimization conditions.  
Since $\partial r_1/\partial \tilde{\tau}$ vanishes, the solution to the 
optimization equation (\ref{F2taut}) is simply
\BE
\label{SP17}
r_1^{\rm opt} =0.
\EE
The second optimization equation, from (\ref{F2tau}), then reduces to 
\BE
\label{SP18}
\at = a (1+ g_1 a),
\EE
and (\ref{g1feq}) gives
\BE
\label{SP19}
\ln(1+ c a) = c \at.
\EE
Eliminating $\at$ between these last two equations gives us 
the optimal $g_1$ in terms of $a$:
\BE
\label{g1result}
g_1^{\rm opt} = \frac{\ln(1+ c a) - c a}{c a^2}.
\EE
%
%
Also, from the  integrated $\beta$-function (``int-$\beta$'') equation (see Appendix B), at 
second order, we have
\BE
\label{SP20}
\tau = \frac{1}{a} + c \ln \frac{c a}{1+ c a}.
\EE
Substituting for $\tau$ and for $g_1$ in Eq.~(\ref{SP16}) and equating to zero, since 
$r_1^{\rm opt}=0$, we find  
\BE
\ln(1+ c a) -(c a)^2 \ln \frac{c a}{1+ c a} = 
c a \left( 2 - a \, \boldsymbol{\sigma}_1(Q) \right),
\EE
which determines the optimized $a$ in terms of the invariant quantities $c$ and 
$\boldsymbol{\sigma}_1(Q)$.  
Substituting back in Eq.~(\ref{g1result}) then fixes $g_1^{\rm opt}$.  The final optimized 
result, from Eq.~(\ref{F2ndord}),  is 
\BE
\label{SP23}
F^{(2)}_{\rm opt} = A ( c a)^g (1+ c a)^{-g(1-g_1^{\rm opt}/c)} .
\EE
Note that the optimization condition $r_1^{\rm opt}=0$ means that $C_{\rm opt}=1$, so that 
all perturbative corrections are effectively exponentiated and re-absorbed into the anomalous 
dimension by the optimization procedure.  As we shall see later, this property holds at any 
order, as first noted by NN \cite{NN1Prog}.

    Also note that while the value of $\at$ (and hence $\tilde{\tau}$) is determined, it is not needed 
to obtain the result for $F^{(2)}_{\rm opt}$.  

\section{RG equations}
\setcounter{equation}{0}

     As discussed above the RS/FS variables are $\tau$, $c_j$, $\tilde{\tau}$, $\tilde{c}_j$ , 
and the $g_i$ coefficients.  We now write down the RG equations expressing the fact that 
the physical quantity $F$ is independent of all these variables.  Symbolically, we have
\BE
\label{RGeqform}
\frac{1}{F} \frac{\partial F}{\partial X} = 0,
\EE
where $X$ stands for any of the set of variables 
$\{ \tau, c_j, \tilde{\tau},  \tilde{c}_j, g_j \} $.

 Recalling the factorized form 
$F = \OO C$ of Eq.~(\ref{Fndef}), and noting that 
$\OO$ is manifestly independent of $\tilde{M}$, we see that 
\BE  
\label{dFdtaut0}
\frac{1}{F} \frac{\partial F}{\partial \tilde{\tau}} = 
\frac{1}{C} \frac{\partial C}{\partial \tilde{\tau}}.
\EE
The same argument applies to the $\tilde{c}_j$ derivatives, since $\OO$, while 
it depends on $a$ and its RS variables $\tau, c_j$, is manifestly independent 
of $\at$ and its RS variables $\tilde{\tau}, \tilde{c}_j$.  Thus, the first two RG 
equations have the familiar form
\BA
\label{dFdtaut1}
\left( \left. \frac{\partial}{\partial \tilde{\tau}} \right|_{\at} + 
\frac{\tilde{\beta}(\at)}{b} \frac{d}{d \at} \right) C \, = 0, 
 & & \quad\quad\quad {\scriptstyle {\mbox{\rm``}} j=1 {\mbox{\rm ''}}} \\ 
& & \nonumber \\
\label{dFdcjt1}
\left( \left. \frac{\partial}{\partial \tilde{c}_j} \right|_{\at} + 
\tilde{\beta}_j(\at) \frac{d}{d \at} \right) C = 0, &  & 
\quad\quad\quad
{\scriptstyle j = 2,3,\ldots,} 
\EA
where the first term collects dependence from the 
$r_i$ coefficients of $C$, while the second term collects the 
compensating dependence via $\at$.  (See Appendix B for the 
definition of the $\beta_j(a)$ functions.)

   The other RG equations all take the form
\BE
\label{dFdX}
\frac{1}{C} \frac{ \partial C}{\partial X} +
\frac{1}{\OO} \frac{ \partial \OO}{\partial X}  = 0,
\EE
where $X$ is any of the variables $\tau, c_j$ or $g_j$.  
The first term only involves dependence via the $r_i$ coefficients -- indeed we 
are tempted to add ``$|_{\at}$'' (meaning ``with $\at$ held constant'') to the notation, 
to match Eqs.~(\ref{dFdtaut1}), (\ref{dFdcjt1}),
but it is unnecessary since $\at$ is manifestly independent of 
$\tau, c_j$ and $g_j$.  
The second term can be evaluated as follows.   In the case $X \to \tau$, 
we may simply use the definition of $ \gammaO$,  Eq.~(\ref{gamOdef}), to get
\BE
\label{dOdtau}
\frac{1}{\OO} \frac{ \partial \OO}{\partial \tau} = \frac{\gammaO}{b}.
\EE
For $X \to c_j$ we can first write
\BE
\label{dOdcj0}
\frac{1}{\OO} \frac{\partial \OO}{\partial c_j} 
=
\left. \frac{1}{\OO} \frac{\partial \OO}{\partial c_j} \right|_{a} + 
\frac{1}{\OO} \frac{d \OO}{d a} 
\frac{\partial a}{\partial c_j},
\EE
and then use  Eq.~(\ref{Adef}) to obtain
\BE
\label{dOdcjA}
\frac{1}{\OO} \frac{\partial \OO}{\partial c_j} = 
\int_0^{a} \! dx \,  \frac{ \gammaO(x)}{\beta(x)^2} \, b x^{j+2} 
+ \frac{\gammaO(a)}{\beta(a)}  \beta_j(a) .
\EE
Although we return to this form later, for the present we follow NN and 
re-write it as  
\BE
\label{dOdcj}
\frac{1}{\OO} \frac{\partial \OO}{\partial c_j} = 
\int_0^{a} \! dx \frac{\beta_j(x)}{\beta(x)} \gammaO^\prime(x),
\EE
where $\gammaO^\prime(x) \equiv d \gammaO/dx$.  The equivalence to 
Eq.~(\ref{dOdcjA}) can be shown by integrating by parts and then using 
the differential equation satisfied by the $\beta_j$ functions 
(see Appendix B).  
Finally, for $X \to g_j$ we find, from  Eq.~(\ref{Adef}), 
\BE
\label{dOdgi}
\frac{1}{\OO} \frac{ \partial \OO}{\partial g_j} = 
- b g \int_0^{a} \! dx \, \frac{ x^{j+1}}{\beta(x)}.
\EE

    Thus, the RG equations, in addition to Eqs.~(\ref{dFdtaut1},\ref{dFdcjt1}), are
\BA
\label{dFdtau2}
\frac{1}{C} \frac{ \partial C}{\partial \tau} + \frac{\gammaO}{b} = 0,
& & 
\quad\quad\quad
{\scriptstyle {\mbox{\rm``}} j=1 {\mbox{\rm ''}}} 
\\
& & \nonumber \\
\label{dFdcj2}
\frac{1}{C} \frac{ \partial C}{\partial c_j} + 
\int_0^{a} \! dx \frac{\beta_j(x)}{\beta(x)} \gammaO^\prime(x),
= 0,
& &  
\quad\quad\quad
{\scriptstyle j=2,3,\ldots,} 
\\
& & \nonumber \\
\label{dFdgi2}
\frac{1}{C} \frac{ \partial C}{\partial g_j} 
- b g \int_0^{a} \! dx \, \frac{ x^{j+1}}{\beta(x)}
= 0,
&  &  
\quad\quad\quad {\scriptstyle j=1,2,\ldots,}
\EA

    As usual, the RG equations determine how the coefficients $r_i$ must depend 
on the RS/FS variables.  We now re-write the RG equations to 
facilitate finding these dependences.   First, we use the series for $\gammaO$ and $C$:
\BE
\gammaO(a) = - b g \sum_{i=0} g_i a^{i+1}, \quad\quad
C = \sum_{i=0} r_i \at^i, 
\EE
with $r_0 \equiv g_0 \equiv 1$.  Second, we convert the $\beta, \beta_j$ functions 
to the $B, B_j$ functions of Appendix B (whose series begin $1+ \ldots$).  A third simplification, 
concerning the lower limit of the $i$ summations, is discussed below.   We 
obtain
\BE
\label{dFdtaut}
\sum_{i=1} \frac{ \partial r_i}{\partial \tilde{\tau}}\, \at^i -  
\at^2 \tilde{B}(\at) \sum_{i=1} i  r_i \at^{i-1} = 0 ,
\EE
\BE
\label{dFdcjt}
\sum_{i=j+1} \frac{\partial r_i}{\partial \tilde{c}_j} \at^i + 
\at^{j+1} \frac{\tilde{B}_j(\at)}{j-1} \sum_{i=1} i r_i \at^{i-1} = 0,
\EE
\BE
\label{dFdtau}
\frac{1}{C} \sum_{i=1} \frac{\partial r_i}{\partial \tau} \,  \at^i 
 - g a \sum_{i=0} g_i  a^{i} = 0 ,
\EE
\BE
\label{dFdcj}
 \frac{1}{C} \sum_{i=j} \frac{\partial r_i}{\partial c_j}\,  \at^i +
\frac{g}{j-1} \int_0^a \! dx \, x^{j-1} 
\frac{B_j(x)}{B(x)}
\sum_{i=0} (i+1) \, g_i x^i
= 0,
\EE
\BE
\label{dFdgj}
\frac{1}{C} \sum_{i=j} \frac{\partial r_i}{\partial g_j} \, \at^i 
+ g \int_0^{a} \! dx \, \frac{x^{j-1}}{B(x)} = 0.
\EE
The $i$ summations of the $\partial r_i/\partial X$ terms inherently begin with $i=1$, 
but in the $c_j$ and $g_j$ equations, where the second term starts only at order $a^j$, 
it is immediately evident that $r_i$ cannot depend on $ c_j$ or $g_j$ for $i <  j$.  
Thus, we may begin those $i$ summations at $i=j$.  
For the $\tilde{c}_j$ equation a stronger result holds, since 
$\partial r_i/\partial \tilde{c}_j$ must vanish for $i=j$ as well as for $i<j$.  
This observation is crucial for the ``exponentiation theorem'' proved in Sect.~5.

    In $(k+1)$-th order all the sums would go up to $i=k$ only and the equations 
would only be satisfied, in an arbitrary RS/FS, up to remainder terms of order 
$a^{k+1}$.  The vanishing of all terms up to and including $a^k$ 
fixes the RS/FS dependence of the $r_i$ coefficients, and leads us to 
identify a set of invariants, $\sigma_j$, as discussed in the next section.

\section{Invariants}
\setcounter{equation}{0}

     The scheme dependences of $r_1$ were already found in Eq.~(\ref{SP15}) 
and led us to the first invariant 
\BE
\boldsymbol{\sigma}_1(Q) =  \tau -g_1 -\frac{r_1}{g}.
\EE
It is $Q$ dependent because $r_1$, when calculated from Feynman diagrams, will 
contain a term $-b g \ln(Q/M)$.  One can view $\boldsymbol{\sigma}_1(Q)$ as 
$b \ln(Q/\Lambda_{F})$, where $\Lambda_F$ is a scale specific to the quantity $F$, 
but related, in an exactly calculable way, to the $\Lambda$ of some 
universal, reference RS.   
The earlier literature used an ``invariant'' $\kappa_1$ given by
\BE
\kappa_1 = r_1 + g g_1 + b g \ln(Q/M).
\EE
It is true that $\kappa_1$ is invariant under changes of FS and renormalization 
scale, with the explicit $g_1$ and $M$ dependences cancelling the implicit $g_1$ 
and $M$ dependences of $r_1$.  Where $\kappa_1$ fails to be invariant is 
under a change of RS that leaves the renormalization scale $M$ unchanged, 
but changes the renormalization prescription, so that $a^\prime =a(1+v_1 a + \ldots)$, 
with some arbitrary $v_1$.  Under such a transformation the $a^g$ factor in $\OO$, 
see Eq.~(\ref{OO1}), becomes $(a^\prime)^g=a^g(1+g v_1 a + \ldots)$, so the 
coefficient $r_1$ must become $r_1^\prime = r_1 - g v_1$ to leave $F=\OO C$ invariant.  
Thus, $\kappa_1^\prime = \kappa_1 - g v_1$.  Since our $\boldsymbol{\sigma}_1(Q)$ is 
\BE
\boldsymbol{\sigma}_1(Q) =  b \ln(Q/\Lambda) - \kappa_1/g ,
\EE
this change in $\kappa_1$ cancels with the change from $\Lambda$ 
to $\Lambda^\prime$, by the Celmaster-Gonsalves \cite{CG} relation. 

     The higher invariants, $\sigma_2, \sigma_3,\ldots$, can be defined to be 
$Q$-independent.  As with the $\rho_j$ invariants, it is convenient to define the 
$\sigma_j$'s so that they reduce to the $\beta$-function coefficients $c_j$ in 
``effective charge'' schemes, defined by the RS/FS choices $g_j=0$, $r_i=0$.    
The invariants, so defined, depend on $\tau$ and $\tilde{\tau}$ only via 
the difference $\tilde{\tau}-\tau$ and have no dependence on $Q$ or $\Lambda$.  

       To find the invariants we will need the conversion between $\at$ and $a$; either 
$\at = a(1+ V_1 a + V_2 a^2 + \ldots)$ or its inverse 
\BE
\label{atoat}
a = \at (1+ \tilde{V}_1 \at  + \tilde{V}_2 \at^2 + \ldots) .
\EE
The $\tilde{V}_i$ coefficients can most easily be found from the relation between 
the $\beta$ functions: $\tilde{\beta}(\at)= (d \at/da) \beta(a)$.  (In fact, the 
calculation mirrors that for the $\rho_i$ invariants in Ref.~\cite{OPTnew}.)   The first 
three coefficients are
\BA
\tilde{V}_1 & =& \tilde{\tau}-\tau , \nonumber \\
\tilde{V}_2 & = & (\tilde{\tau}-\tau)^2 + c (\tilde{\tau}-\tau) - (\tilde{c}_2-c_2), 
\label{Vtiresults} \\
\tilde{V}_3 & = & (\tilde{\tau}-\tau)^3 + \frac{5}{2} c (\tilde{\tau}-\tau)^2 + 
 (-2 \tilde{c}_2+3 c_2)(\tilde{\tau}-\tau) -\frac{1}{2}(\tilde{c}_3-c_3).
\nonumber
\EA
Note that the $\tilde{V}_i$'s do {\it not} only involve differences $c_j-\tilde{c}_j$.  
It {\it is} true, though, that the $V_i$ coefficients of the inverse relationship are 
obtained by exchanging all plain and tilde variables.  

We now turn to a calculation of the invariant $\sigma_2$.
Expanding Eqs.~(\ref{dFdtaut}--\ref{dFdgj}) in powers of $a$ and $\at$ and using 
the above result for $\tilde{V}_1$,  we can extract the self-consistency conditions.
From the lowest-order terms we recover Eqs.~(\ref{SP15}) for $r_1$'s derivatives, 
plus confirmation that $r_1$ does not depend on the other RS/FS variables ($c_2$, 
$\tilde{c}_2$, $g_2$).  From the next-order terms we find 
\BA
\frac{\partial r_2}{\partial \tilde{\tau}} =  r_1, & \quad\quad & 
\frac{\partial r_2}{\partial \tau} = g \left( r_1 +g_1 + \tilde{\tau} - \tau \right), 
\nonumber \\
\frac{\partial r_2}{\partial \tilde{c}_2} =0, &  & 
\frac{\partial r_2}{\partial c_2} =-\frac{g}{2} ,
\label{r2derivs}
\\
\frac{\partial r_2}{\partial g_1} = -g \left(r_1- \frac{c}{2} + \tilde{\tau} - \tau \right), 
& &
\frac{\partial r_2}{\partial g_2} = - \frac{g}{2}. 
\nonumber
\EA
Integrating each of these equations individually is easy, but combining the results 
consistently is a little tricky.  However, it is straightforward to check our 
result that $r_2$ has the form:
\BE
r_2= \frac{1}{2} \left( -g c_2 + g g_1 c+ g g_1^2 - g g_2 + 2 g_1 r_1 + 
r_1^2 + \frac{r_1^2}{g} + 2 r_1 (\tilde{\tau}-\tau) \right) +\mbox{ {\rm const.}},
\EE
where the constant is independent of all the RS/FS variables.  The constant can be 
conveniently written as 
$\frac{g}{2} \sigma_2$ so that the invariant $\sigma_2$ is given by
\BE
\sigma_2 = c_2 + g_2 - g_1 c - g_1^2 + \frac{2 r_2}{g} - 2 g_1 \frac{r_1}{g} 
- \frac{r_1^2}{g^2} (1+g) - \frac{2 r_1}{g}(\tilde{\tau}-\tau).
\EE

      An easier and more systematic way to calculate the $\sigma_i$ invariants is to 
find them as the $\rho_i$ invariants associated with the physical quantity
\BE
\label{calDdef}
{\cal D} \equiv \frac{Q}{F}\frac{dF}{dQ}.
\EE
The perturbation series for ${\cal D}$ can be found in terms of the $C$ and $\gammaO$ 
series in various ways.  Perhaps the simplest is the following.  First, note that all the 
$Q$ dependence of $F$ resides in the $r_i$ coefficients of $C$.  
For dimensional reasons such $Q$ dependence can come only via the ratios 
$Q/M$ and $Q/\tilde{M}$.  Thus,
\BE
{\cal D} = \frac{Q}{C}\frac{dC}{dQ} = 
- \frac{1}{C} \left( M \frac{dC}{dM} + 
\left. \tilde{M} \frac{\partial C}{\partial \tilde{M}}\right|_{\at} \right).
\EE
The $M$ dependence of $C$ must cancel out with that of $\OO$ in 
the product $F=\OO C$, so that
\BE
\frac{M}{C} \frac{dC}{dM} = - \frac{M}{\OO} \frac{d\OO}{dM} = - \gammaO,
\EE
while $C$ is independent of $\tilde{M}$, so that 
\BE
0 = \tilde{M} \frac{d C}{d \tilde{M}} = 
\left. \tilde{M} \frac{\partial C}{\partial \tilde{M}} \right|_{\at}
+ \tilde{\beta}(\at) \frac{dC}{d\at}.
\EE
From these observations we see that 
\BE
\label{calDform}
{\cal D} =  \gammaO + \frac{\tilde{\beta}(\at)}{C} \frac{dC}{d\at}.
\EE
Thus, ${\cal D}$ is, in a sense, a ``physicalized'' version of $\gammaO$.  

    Substituting in the above formula we find
\BE
{\cal D} = - b g a(1+g_1 a + g_2 a^2 + \ldots) + 
(-b \at^2)(1+ c \at + \ldots) 
\frac{(r_1 + 2 r_2 \at + \ldots)}{(1+ r_1 \at + \ldots)}.
\EE
We could now expand out in terms of $\at$, converting $a$ 
to $\at$ using Eq.~(\ref{atoat}).  Alternatively, we can eliminate 
$\at$ and find the series expansion in terms of $a$.  The results are more compact 
in the $a$ scheme:
\BE
{\cal D} = - b g a(1+ r_1^{{\cal D}} a + 
r_2^{{\cal D}} a^2 + \ldots),
\EE
with
\BA 
r_1^{{\cal D}} & = & g_1 + r_1/g, 
\\
r_2^{{\cal D}} & = & g_2 +  
\frac{1}{g} \left( 2 r_2 + c r_1 - r_1^2 - 2 r_1(\tilde{\tau}-\tau) \right),
\EA
and so on.  Note that these coefficients are independent of the FS and 
independent of the tilde RS variables, with the explicit $g_i$ and 
$\tilde{\tau}, \tilde{c}_j$ dependences exactly cancelling with the 
implicit dependences from the $r_i$ coefficients; see 
Eqs.~(\ref{SP15}), (\ref{r2derivs}).  Thus, the $r^{{\cal D}}_i$ 
coefficients only depend, in the usual way, on the RS variables 
$\tau, c_j$ associated with $a$.  

     As usual, we can construct the $\rho_j$ invariants for the quantity 
${\cal D}$:
\BA
\boldsymbol{\rho}_1^{{\cal D}}(Q) & = & \tau - r_1^{{\cal D}} ,
\\
\rho_2^{{\cal D}} & = & c_2 + r _2^{{\cal D}} - 
c r_1^{{\cal D}} -(r_1^{{\cal D}})^2 ,
\EA
and these coincide with the $\sigma$'s.  Indeed, it is easy to see 
that the ``effective-charge-type'' RS/FS used in the definition of the $\sigma$'s
corresponds to the usual effective-charge scheme for ${\cal D}$, so the 
equivalence of $\rho_j^{{\cal D}}$ to $\sigma_j$ is true for all $j$.  

The calculation can be straightforwardly extended to higher orders.  
Defining
\BE
\label{Deltadef}
%
%
\Delta \equiv \tilde{\tau}-\tau = b \ln(\tilde{M}/M), \quad\quad  s_i \equiv \frac{r_i}{g},
\EE
the first three invariants are 
\BA
& & \quad \quad 
\boldsymbol{\sigma}_1(Q)  =  \tau -g_1 - s_1,
\\ 
\sigma_2 & = &  c_2 + g_2 - g_1 c - g_1^2 + 2 s_2 - 2 g_1 s_1
- s_1^2 (1+g) - 2 s_1 \Delta ,
\\
\sigma_3 & = & c_3 + c g_1^2 + 4 g_1^3 - 6 g_1 g_2 + 2 g_3 -2 c_2 g_1
+ 6(\tilde{c}_2-c_2) s_1 
- 4 c g_1 s_1 
\nonumber  \\
& & 
 + 12 g_1^2 s_1 - 6 g_2 s_1 - 
 5 c s_1^2 - 2 c g s_1^2 + 12 g_1 s_1^2 + 6 g g_1 s_1^2 + 4 s_1^3 + 
 6 g s_1^3
\nonumber  \\
& & 
 + 2 g^2 s_1^3 + 4 c s_2 - 12 g_1 s_2 - 
 12 s_1 s_2 - 6 g s_1 s_2 + 6 s_3
\nonumber  \\
& & 
+ (12 g_1 s_1 - 10 c s_1  + 12 s_1^2 + 6 g s_1^2 - 12 s_2) \Delta + 6 s_1 \Delta^2 .
\EA
Using these formulas the values of the invariants can be found from Feynman-diagram 
calculations performed in any convenient RS/FS.

\section{The exponentiation theorem}
\setcounter{equation}{0}

      The $(k+1)$-th order approximation is defined by truncating the series for 
$C$, $\gammaO$, $B$, and $\tilde{B}$.  The resulting approximant, in general, will have a 
residual RS/FS dependence that is formally of order $a^{k+1}$.  The optimization conditions 
correspond to requiring the RG equations to be exactly satisfied, with no remainder.  
(To avoid notational clutter, we leave it understood that, henceforth, any RS/FS-dependent 
symbol ($a, \at, r_i,$ etc.) stands for the optimized value of that quantity.)   

       At second order we saw that the $\tilde{\tau}$ optimization equation gave $r_1=0$.  In 
third order $(k=2$)  the $\tilde{\tau}$ equation (\ref{dFdtaut}), in which  
$\partial r_2/\partial \tilde{\tau} = r_1$, reduces to
\BE
(1+ c \at + \tilde{c}_2 \at^2)(r_1+2 r_2 \at) - r_1 =0.
\EE
Also, the $\tilde{c}_2$ equation (\ref{dFdcjt}), in which the $\tilde{B}_2(\at)$ factor cancels 
out because $\partial r_2/\partial \tilde{c}_2=0$, becomes just
\BE
\label{lasteq}
r_1+ 2 r_2 \at =0.
\EE
Substituting this back into the previous equation gives $r_1=0$.  Substituting $r_1=0$ back 
into Eq.~(\ref{lasteq}) then gives $r_2=0$.  The result generalizes to all orders, as first noted by NN.  

\vspace*{3mm}

{\bf Theorem} (Nakkagawa and Ni\'egawa \cite{NN1Prog})

The solution to the $\tilde{\tau}$ and $\tilde{c}_j$ optimization equations is
\BE
 r_1=r_2=\ldots=r_k=0.
\EE
Thus, $C=1$ in the optimal scheme, so that all perturbative corrections are effectively 
exponentiated and re-absorbed into the anomalous dimension $\gammaO$.

\vspace*{3mm}

{\bf Proof:}  The $\tilde{c}_j$ optimization equation follows from Eq.~(\ref{dFdcjt}):
\BE
\label{dFdcjtcopy}
\sum_{i=j+1}^k \frac{\partial r_i}{\partial \tilde{c}_j} \at^i + 
\at^{j+1} \frac{\tilde{B}_j(\at)}{j-1} \frac{d C}{d \at}  = 0,
\EE
where $d C/d \at = \sum_{i=1}^k i r_i \at^{i-1}$.  
Recall that all terms up to and including $\at^k$ must cancel in any RS, thus 
determining $\partial r_i/\partial \tilde{c}_j$.  By starting the sum at $i=j+1$ we have already 
used the fact that $\partial r_i/\partial \tilde{c}_j$ must vanish for $i<j$ {\it and for} $i=j$, as 
noted at the end of Sect.~3.  

   We begin by considering the case $j=k$.  The first term vanishes, as there are no terms in 
the sum, so we find that in the optimal scheme
\BE
\frac{d C}{d \at} =0.
\EE 
Next, consider the case $j=k-1$.   In any scheme, cancellation of 
the $\at^k$ terms requires
\BE
 \frac{\partial r_k}{\partial \tilde{c}_{k-1}} = - \frac{r_1}{k-2}.
\EE
In the optimal scheme the left-hand side must vanish, since $d C/d \at$ vanishes in the 
optimization equation (\ref{dFdcjtcopy}).  Thus, in 
the optimal scheme, $r_1=0$. 
Proceeding to the case $j=k-2$ we can find $\partial r_k/\partial \tilde{c}_{k-2}$ as a 
sum of $r_1 c$ and $r_2$ terms.  In the optimal scheme this must vanish, and since we 
already have $r_1=0$, we now find that $r_2=0$, too.  We may then proceed to successively 
lower $j$ cases to see that other $r_i$'s vanish.  Finally, we reach $j=1$, where we are dealing 
with the $\tilde{\tau}$ equation, which gives us $r_{k-1}=0$.  Substituting back into 
$dC/d \at= \sum_{i=1}^k i r_i \at^{i-1}=0$ then shows that $r_k=0$.

\section{The optimization equations}
\setcounter{equation}{0}

         The fact that $C=1$ in the optimal scheme allows us to simplify the remaining 
optimization equations, which follow from Eqs.~(\ref{dFdtau}--\ref{dFdgj}) with the 
$i$ summations truncated at $i=k$.  

Also, recalling that the $B_j(a)$ functions are related 
to the $I_j(a)$ integrals, one sees that the $c_j$ equation involves  
\BE
\label{Iij}
I_{j,i}(a)  \equiv  (i+1) \int_0^a \! dx \,  x^i I_j(x).
\EE
This can be simplified by interchanging the order of the two integrations:
\BA
I_{j,i}(a) &=& (i+1) \int_0^a \! dx \, x^i \int_0^x \! dy  \frac{y^{j-2}}{B(y)^2} 
\nonumber \\
&=& \int_0^a \! dy \, \frac{y^{j-2}}{B(y)^2} \int_y^a \! dx \, (i+1) x^i 
\nonumber \\
&=&  \int_0^a \! dy \,  \frac{y^{j-2}}{B(y)^2} \left( a^{i+1} - y^{i+1} \right) ,
\EA 
to give
\BE
I_{j,i}(a) = a^{i+1} I_j(a) -  I_{i+j+1}(a), 
\EE
which corresponds to going back to the form in Eq.~(\ref{dOdcjA}) for 
$\frac{1}{\OO} \frac{\partial \OO}{\partial c_j}$.   Also note that 
the $g_j$ optimization equations involve a related set of integrals
\BE
\label{Jninteg}
J_j(a) \equiv \int_0^a \! dx \frac{x^{j-2}}{B(x)}.
\EE

    Thus, the $\tau$, $c_j$, and $g_j$ optimization equations can be written as
\BE
\label{tauopteq}
\sum_{i=1}^k \frac{\partial r_i}{\partial \tau} \,  \at^i 
 - g a \sum_{i=0}^k g_i  a^{i} = 0 , 
\quad\quad\quad {\scriptstyle {\mbox{\rm ``}} j=1 {\mbox{\rm ''}}}
\EE
\BE
\label{cjopteq}
\sum_{i=j}^k \frac{\partial r_i}{\partial c_j}\,  \at^i +
g \sum_{i=0}^k g_i  I_{j,i}(a)
= 0,
\quad\quad\quad 
{\scriptstyle j=2,\ldots,k}
\EE
\BE
\label{gjopteq}
\sum_{i=j}^k \frac{\partial r_i}{\partial g_j} \, \at^i 
+ g J_{j+1}(a) = 0.
\quad\quad\quad\quad\,\,
{\scriptstyle j=1,\ldots,k}
\EE

In each of these equations the first term is a polynomial in $\at$ that must 
precisely cancel out the terms up to and including $\at^k$ present in the second term, 
if it were expanded out in a power series in $\at$.  In Ref~\cite{OPTnew} we 
used the notation 
$\mathbb{T}_n [G(a)]$ to mean ``truncate the series for 
$G(a)=G_0+G_1 a+\ldots$ immediately after the $a^n$ term''  
(i.e.,  $\mathbb{T}_n [G(a)] \equiv G_0 + G_1 a + \ldots + G_n a^n$).  
Here we will need $\tilde{\mathbb{T}}_n$ as the equivalent operation in 
the expansion parameter $\at$.  Thus, we may re-write the equations 
(swapping the order of the two terms and dividing out a $g$ factor) as
\BE
\label{tauopteq1}
 a \sum_{i=0}^k g_i  a^{i} - \tilde{\mathbb{T}}_k[ \, a \sum_{i=0}^k g_i  a^{i} \, ] = 0,
\quad\quad\quad\quad  {\scriptstyle {\mbox{\rm ``}} j=1 {\mbox{\rm ''}}}
\EE
\BE
\label{cjopteq1}
\sum_{i=0}^k g_i  I_{j,i}(a) - \tilde{\mathbb{T}}_k[\, \sum_{i=0}^k g_i  I_{j,i}(a) \, ] = 0,
\quad\quad\quad 
{\scriptstyle j=2,\ldots,k}
\EE
\BE
\label{gjopteq1}
 J_{j+1}(a)  - \tilde{\mathbb{T}}_k[ \, J_{j+1}(a) \,] = 0,
\quad\quad\quad\quad\quad\quad
{\scriptstyle j=1,\ldots,k}
\EE
However, note that the arguments of the $ \tilde{\mathbb{T}}_k$'s are all functions 
of $a$, rather than $\at$, so it is best to think of the $ \tilde{\mathbb{T}}_k[G]$ operation 
in three stages (i) expand $G$ as series in $a$ up to $a^k$, (ii) convert $a$ to $\at$ using 
Eq.~(\ref{atoat}), and 
(iii) re-expand as a series in $\at$, and truncate after the $\at^k$ term.  

   A further simplification results from the realization that, since $C=1$, we do 
not need to know the optimized value of $\at$; nor do we need to know the $\tilde{c}_j$'s 
or $\tilde{\tau}$: they do not enter into the optimized result for $F$, which just involves 
evaluating $\OO$ in the optimal scheme.  Thus, what we need to do is to 
take combinations of the optimization equations in which $\at$ and the $\tilde{V}_i$'s 
cancel out.  
From the resulting equation combinations we can solve for the $g_j$ coefficients in 
terms of the  ``principal variables'' $a, c_2, \ldots c_k$.  (Note that the $I$ and $J$ 
integrals are functions of these principal variables.)  Finally, we can use the invariants, 
$\sigma_i$ and $\boldsymbol{\sigma}_1(Q)$, and the int-$\beta$ equation to determine 
the optimized result.  Note that when $r_i$=0 the $\sigma_j$'s have exactly the same 
form as the usual $\rho_j$ invariants with $g_i$'s in place of $r_i$'s.  


       In the next section we illustrate the above observations in the case of third 
order.

\section{Third-order approximation}
\setcounter{equation}{0}

       In third order ($k=2$) we have four remaining optimization equations, in the 
variables $\tau$, $c_2$, $g_1$, and $g_2$.  From Eqs.(\ref{tauopteq1})--(\ref{gjopteq1}) 
these are
\BA
a(1+g_1 a+ g_2 a^2) - \at - (g_1 + \tilde{V}_1)\at^2 = 0, 
& \quad\quad\quad & {\scriptstyle (\tau)}
\\
 I_{2,0}+g_1 I_{2,1} +g_2 I_{2,2} - \frac{1}{2}\at^2 =0,
& & {\scriptstyle (c_2)}
\\
J_2 - \at - \left( -\frac{c}{2} + \tilde{V}_1 \right) \at^2 = 0,
& & {\scriptstyle (g_1)}
\\
J_3 - \frac{1}{2}\at^2 = 0.  
& & {\scriptstyle (g_2)}
\EA
Taking the $g_1$ equation minus the $\tau$ equation cancels the $\at$ terms 
and, not coincidentally, the $\tilde{V}_1$ terms, leaving
\BE
J_2 - a(1+g_1 a+ g_2 a^2) + \left( \frac{c}{2} + g_1 \right) \at^2 = 0.
\EE
An $\at^2$ term remains, but we can substitute from the $g_2$ equation to 
obtain
\BE
\label{k2combeq1}
J_2 + (c + 2 g_1) J_3 - a(1+g_1 a+ g_2 a^2) = 0.
\EE
Taking the $g_2$ equation minus the $c_2$ equation cancels the $\at^2$ terms, 
giving
\BE
\label{k2combeq2}
J_3 - \left(  I_{2,0} +g_1 I_{2,1} +g_2 I_{2,2} \right) =0.
\EE
We may solve these last two equations for $g_1, g_2$ in terms of the principal 
variables $a, c_2$. 


   From the four original equations we have extracted just two equations that give us 
the $g_1, g_2$ coefficients that we need.  There are effectively two other 
equations that we can just ignore; they would determine $\at$ and $\tilde{V}_1$ 
(which gives $\tilde{\tau}$ and, combined with the int-$\tilde{\beta}$ equation of 
the tilde scheme, would then fix $\tilde{c}_2$), but we have no need to obtain 
values for these variables.  

    To relate the principal variables to $Q$ and the invariants, we substitute the 
optimal-scheme quantities into the expressions for 
$\sigma_2$ and $\boldsymbol{\sigma}_1(Q)$, combining the latter with the int-$\beta$ 
equation to eliminate $\tau$.  In the optimal scheme, since $r_i=0$, the formula 
for $\sigma_2$ reduces to
\BE
\sigma_2 = c_2 + g_2 - g_1 c -g_1^2,
\EE
which is the familiar form of a $\rho_2$ invariant, but with $g_i$'s as the coefficients.  
Similarly, in the optimal scheme
\BE
\boldsymbol{\sigma}_1(Q) = \tau -g_1 = K^{(3)}(a) - g_1,
\EE
where $K^{(3)}(a)$ is the third-order approximation to the $K(a)$ function 
of the int-$\beta$ equation.  

\section{A simpler approach}
\setcounter{equation}{0}


    In fact, there is a simpler approach that allows us to get directly to the 
equations determining the optimal $g_i$'s.  Consider the physical quantity 
${\cal D}$ defined in Eq.~(\ref{calDdef}), which we showed is given by 
Eq.~(\ref{calDform}), so that ${\cal D}=\gammaO$ when $C=1$.  That 
suggests that we consider $F$ in the form:
\BE
\label{Fnewform}
F = A  \exp \int_{[0]}^{a} dx \,
\frac{{\cal D}(x)}{\beta(x)},
\EE
where ``$[0]$' is a shorthand for the same ``lower limit of $0$ with subtraction of 
the suitable infinite scheme-independent constant,'' as in Eq.~(\ref{Adef}).  
Formally, this expression for $F$ is valid quite generally, and is independent of the 
RS used, so it satisfies RG equations saying that the total dependences on $\tau$ 
and $c_j$ all vanish.  What we 
are doing in RS/FS optimization is equivalent to a normal RS optimization applied to 
$F$, except that the approximants being optimized are not truncations of the 
perturbation series for $F$, but are approximants formed by truncating 
the perturbation series for ${\cal D}$ and $\beta$.  That is, the $(k+1)$-th 
approximant to $F$ is given by substituting 
\BE
 {\cal D}(x) = \sum_{i=0}^k r_i^{{\cal D}} x^{i+1}, \quad\quad 
 \beta(x) = - b x^2 \sum_{j=0}^k c_j x^k
\EE
into Eq.~(\ref{Fnewform}).  The optimization equations follow from requiring 
the $\tau$ and $c_j$ derivatives to vanish. (Note that when we take such 
derivatives the infinite constant plays no role and the ``$[0]$'' lower limit can 
safely be replaced by $0$, since the resulting integrals converge.)  For $\tau$ we have 
\BA
0=
\frac{1}{F} \frac{\partial F}{\partial \tau} & = &
\frac{\partial a}{\partial \tau} \frac{{\cal D}(a)}{\beta(a)} + 
\int_0^a dx \left. \frac{ \partial {\cal D}}{\partial \tau}\right|_x \frac{1}{\beta(x)}
\nonumber \\
\label{newtaueq}
& = & \frac{1}{b} \left( {\cal D}(a) - 
\sum_{i=1}^k \frac{ \partial r_i^{{\cal D}}}{\partial \tau} J_{i+1} \right),
\EA
while for $c_j$
\BA
0=
 \frac{1}{F} \frac{\partial F}{\partial c_j} & = & 
\frac{\partial a}{\partial c_j} \frac{{\cal D}(a)}{\beta(a)} + 
\int_0^a dx \left(  \left. \frac{ \partial {\cal D}}{\partial c_j}\right|_x \frac{1}{\beta(x)}
+ \frac{{\cal D}(x)}{\beta(x)^2} b x^{j+2} \right)
\nonumber \\
\label{newcjeq}
& = & 
- \frac{1}{b} \left( - {\cal D}(a) I_j - 
\sum_{i=j}^k  \frac{ \partial r_i^{{\cal D}}}{\partial c_j} J_{i+1} 
+ \sum_{i=0}^k r_i^{{\cal D}} I_{i+j+1} 
\right).
\EA
Substituting the series form for ${\cal D}(a)$ leads to 
\BA
\label{newtaueqB}
- \sum_{i=1}^k  \frac{ \partial r_i^{{\cal D}}}{\partial \tau} J_{i+1}
+ \sum_{i=0}^k r_i^{{\cal D}} a^{i+1} 
& = & 0,
\\
\label{newcjeqB}
\sum_{i=j}^k  \frac{ \partial r_i^{{\cal D}}}{\partial c_j} J_{i+1} 
+ \sum_{i=0}^k r_i^{{\cal D}} I_{j,i}
& = & 0,
\EA
where $I_{j,i}(a) = a^{i+1} I_j(a) -  I_{i+j+1}(a)$ arises from the first and 
third terms of Eq.~(\ref{newcjeq}).

    The derivatives $\partial r_i^{{\cal D}}/\partial \tau$ and 
$\partial r_i^{{\cal D}}/\partial c_j$ are the usual RS dependences 
of perturbative coefficients \cite{OPT,OPTnew}, and can be quickly found 
from the expressions for the $\rho_i^{{\cal D}}$ invariants.  Thus,
\BE
 \frac{ \partial r_1^{{\cal D}}}{\partial \tau} =1, 
\quad\quad
 \frac{ \partial r_2^{{\cal D}}}{\partial \tau} = c + 2 r_1^{{\cal D}},
\quad\quad
 \frac{ \partial r_2^{{\cal D}}}{\partial c_2} = -1.
\EE
Using these results, and recalling that in the FS/RS optimal scheme the 
optimized $r_i^{{\cal D}}$'s equal the optimized $g_i$'s, 
the reader can quickly check that at 3rd order ($k=2$) 
Eqs.~(\ref{newtaueqB}) and (\ref{newcjeqB}) lead directly to Eqs.~(\ref{k2combeq1}) 
and (\ref{k2combeq2}).  

      At 4th order ($k=3$) the $\tau, c_2, c_3$ equations reduce to
\BE
J_2 + (c+2 g_1) J_3 + (c_2 + 2 c g_1 + 3 g_2)J_4 
- a(1+g_1 a+g_2 a^2 + g_3 a^3) =0,
\EE
\BE
J_3+2 g_1 J_4 -
(I_{2,0} + g_1 I_{2,1} + g_2 I_{2,2} + g_3 I_{2,3}) =0,
\EE
\BE
\frac{1}{2} J_4  - 
(I_{3,0} + g_1 I_{3,1} + g_2 I_{3,2} + g_3 I_{3,3}) =0.
\EE
We have explicitly checked that these are indeed the equations one would 
obtain from appropriate combinations of 
Eqs.~(\ref{tauopteq1}), (\ref{cjopteq1}), (\ref{gjopteq1}).

\section{Conclusions and outlook}
\setcounter{equation}{0}

       The optimization approach to the problem of RS/FS dependence is now, we believe, on a 
firm footing.  It is far less daunting than it might appear at first sight. There are  
$3k$ scheme variables at $(k+1)$-th order and $k$ coefficients, $r_i$.   
However, $k$ of the optimization equations lead to $r_1=\ldots=r_k=0$, so that $C=1$; 
another $k$ variables ($\tilde{\tau}, \tilde{c}_2,\ldots,\tilde{c}_k$)  then need not be solved 
for.  That leaves $k$ combinations of 
optimization equations that can be solved for $g_1,\ldots,g_k$ in terms of the ``principal 
variables'' $a, c_2,\ldots,c_k$.  In fact, these equations can be obtained more directly by 
the approach in the last section.  By substituting in the expressions for the invariants, 
one can then solve for all the needed quantities.  The last step will require an iterative algorithm, 
as in ordinary optimization \cite{OPTnew}.

       Our results have applications to various quantities, such as charmonium decays to 
hadrons, $B$ decays to charmonium, or Higgs boson decay to hadrons:  These quantities 
have a factorized form involving the wavefunction at the origin or, in the last case, the quark 
masses.  For applications involving parton distribution functions and fragmentation functions there is 
more work to be done.  We have only considered the non-singlet case; the flavour-singlet case
involves matrices describing quark-gluon mixing.  Also, our analysis has used the language of 
structure-function moments, which is convenient theoretically since it reduces a convolution 
integral to a simple product.  However, phenomenologically, it seems preferable to deal directly 
with the parton distributions using parton-evolution (DGLAP) equations.  It would be valuable 
to see if our moments-based approach can be reformulated in that language and put into practice.  

       We end with a plea to recognize of the importance of this effort.  When QCD was 
young, the use of phenomenological, {\it ad hoc} choices was excusable, perhaps even 
necessary to make progress.  Now that the theory is mature we cannot go on using arbitrary 
renormalization prescriptions and blind guesses at the ``right'' renormalization and factorization 
scales (which don't even exist, since it is only the ratios of $M$ and $\tilde{M}$ to the 
prescription-dependent $\Lambda$ that matter).  If ``precision QCD'' is to be a valid scientific 
enterprise, it must be based on a systematic treatment of RS/FS ambiguities, with a respect for 
RG invariance at its core.

\newpage

\section*{Appendix A:  Discussion of the work of NN}
\renewcommand{\theequation}{A.\arabic{equation}}
\setcounter{equation}{0}

   In this appendix we critique the work of Nakkagawa and Ni\'egawa (NN) \cite{NN1}-\cite{NN3} 
and outline why, nevertheless, their optimization equations are equivalent to ours.  Note that 
their ``$\mu$'' corresponds to our $\tilde{M}$ (and their ``$b$'' is the opposite sign to ours).  
Their $\at$ is the same as ours, but their $a$ is somehow supposed to explicitly depend on 
{\it both} $M$ and $\tilde{M}$.  They write $a=a(\mu, \xi)$ where $\xi=M/\mu$.  It is never 
clear quite how this object is defined.  Because of its supposed dependence on {\it two} scales, 
NN associate it with {\it two} $\beta$ functions, whose coefficients are supposed to depend on 
$\xi$.   We find this rather odd; it might not be wrong, but it certainly creates difficulties without 
gaining any generality.  In our approach the couplant $a$ is a normal couplant, 
with a renormalization scale $M$, in a RS labelled by $\tau\equiv b \ln(M/\Lambda), c_2, c_3, \ldots$.  
This RS is distinct from, and independent of, the tilde RS used for $\at$, whose scale is 
$\tilde{M}$ and whose scheme labels are $\tilde{\tau}, \tilde{c}_2, \tilde{c}_3, \ldots$.  
Along with FS labels $g_1, g_2, \ldots$ these form the complete set of RS/FS labels, and 
variation of any one label, in a partial derivative, is made holding the other labels constant.  
Thus, there is no question of $c_j$'s ``depending'' on $M$ or $\tilde{M}$ or their ratio.  

    For NN the integration of their two $\beta$-function equations for ``$a(\mu,\xi)$'' is 
problematic \cite{NN1Prog,NN2}, because of a dependence on the integration path.  Later \cite{NN3} 
they claimed to have resolved this problem, and made the $\xi$ dependence of their $c_j$'s go away.  
In our view, this dependence and the integration-path problem should never have been there in the 
first place!  

      NN's analysis involves a somewhat mysterious variable $\Phi$, which it seems must 
actually be, in their notation, $b \ln(M/\mu)$.  In our notation that means 
$\Phi=-b\ln(M/\tilde{M}) = \tilde{\tau}-\tau$.  Provided that we make this 
identification, we find that their equations (Eqs.~(18a-e) of Ref.~\cite{NN1Prog}) 
are equivalent to ours.  Apart from straightforward 
conversion of notation we need to recognize that they work with variables $\mu$ and $\Phi$, 
etc., while we work with $\tilde{M}=\mu$ and $M$ (related to $\tilde{\tau}$ and $\tau$, 
respectively).  Thus their $\partial /\partial \Phi$ is at constant $\mu$ and coincides with our 
$-(1/b) M \partial /\partial M = -\partial/\partial \tau$:  However, their 
$\mu \partial /\partial \mu$ is at constant $\Phi$ and so corresponds to our 
$\tilde{M} \partial/\partial \tilde{M} + M \partial/\partial M = b (\partial/\partial \tilde{\tau} 
+\partial/\partial \tau$).  Hence, their optimization equation associated with $\mu$ 
is a sum of our $\tilde{\tau}$ and $\tau$ optimization equations.  

   Notwithstanding our criticisms, NN deserve praise for arriving at the correct optimization 
equations, and they were correct to criticize Refs.~\cite{Politzer,StePol}'s formulation 
as insufficiently general.  The applications of their results, pursued with Yokota \cite{NNY}, 
are valid and important.  In particular, they show how optimization naturally resolves the 
issue that, in a na\"ively fixed scheme, the perturbative coefficients for the $n$th moment 
would grow like $\ln n^2$.

\section*{Appendix B:  $\beta(a)$ and $\beta_j(a)$ functions}
\renewcommand{\theequation}{B.\arabic{equation}}
\setcounter{equation}{0}

    For the reader's convenience we list here some key formulas from 
Refs.~\cite{OPT,OPTnew}.  
The integrated form of the $\beta$-function equation, referred to as the 
``int-$\beta$'' equation, is
\BE
\label{intbeta0}
\tau \equiv b \ln(M/\Lambda) = 
\lim_{\delta \to 0} \left( \int_\delta^{a} 
\frac{d x}{ \beta(x)} + {\cal C(\delta)} \right) 
\equiv K(a), 
\EE
with
\BE
\label{Cdeltadef}
{\cal C}(\delta) \equiv \int_\delta^{\infty} 
\frac{d x}{ b x^2(1 + c x)}.  
\EE

       The $\beta_j$ functions, defined as $\partial a/\partial c_j$, are given by
\BE
\label{betaj}
\beta_j(a) = -b \beta(a) \int_0^a dx \; \frac{x^{j+2}}{\beta(x)^2}.
\EE
Their series  expansions begin at order $a^{j+1}$ so it is convenient to define $B_j(a)$ 
functions which begin $1+O(a)$:
\BE
B_j(a) \equiv \frac{(j-1)}{a^{j+1}} \beta_j(a).
\EE
For $j=1$ it is natural to define 
\BE
B_1(a) \equiv B(a) \equiv \frac{\beta(a)}{-b a^2} = 1+ c a+ c_2 a^2 + \ldots
 =  \sum_{i=0}^{\infty} c_i a^i,
\EE
with the convention that $c_0 \equiv 1$ and $c_1 \equiv c$.  
Equation (\ref{betaj}) can then be re-written as
\BE
\label{Bj}
B_j(a) = \frac{(j-1)}{a^{j-1}} B(a) I_j(a),
\EE
where 
\BE
\label{Ij}
I_j(a) \equiv \int_0^a dx \, \frac{x^{j-2}}{B(x)^2}.
\EE
(Note that this formula for $B_j(a)$ even holds for $j=1$ if the r.h.s. is 
interpreted as the limit $j \to 1$ from above.)

   Differentiating Eq.~(\ref{betaj}) leads to 
\BE
\label{diffeq}
\beta_j^{\prime}(a) \beta(a) - \beta^{\prime}(a) \beta_j(a) = - b a^{j+2},
\EE
where here the prime indicates differentiation with respect to $a$, regarding 
the coefficients $c_j$ as fixed.


\newpage


\begin{thebibliography} {99}

\bibitem{OPT} 
  P. M. Stevenson, Phys. Rev. D {\bf 23}, 2916 (1981).

\bibitem{Politzer} 
  H. D. Politzer, Nucl. Phys. B {\bf 194}, 493 (1982).

\bibitem{StePol}
  P. M. Stevenson and H. D. Politzer, Nucl. Phys. B {\bf 277}, 758 (1986).

\bibitem{NN1}
  H. Nakkagawa and A. Ni\'egawa, Phys. Lett. B {\bf 119}, 415 (1982).
 
\bibitem{NN1Prog}
 H. Nakkagawa and A. Ni\'egawa,  Prog. Theor. Phys. {\bf 70}, 511 (1983).

\bibitem{NN2}
  H. Nakkagawa and A. Ni\'egawa, Prog. Theor. Phys. {\bf 71}, 339 (1984).

\bibitem{NN3}
  H. Nakkagawa and A. Ni\'egawa, Prog. Theor. Phys. {\bf 71}, 816 (1984).

\bibitem{OPTnew}
  P. M. Stevenson, Nucl. Phys. B {\bf 868}, 38 (2013).

\bibitem{CG} 
  W. Celmaster and R. J. Gonsalves, Phys. Rev. D {\bf 20}, 1420 (1979). 

\bibitem{NNY}
  H. Nakkagawa, A. Ni\'egawa, and H. Yokota, Phys. Rev. D {\bf 34}, 244 (1986).


\end{thebibliography}
\end{document}